\documentclass[preprint2]{aastex}
\usepackage{spr-astr-addons}
\usepackage{graphicx}

\newcommand{\bbv}[1]{\mbox{\boldmath$#1$}}
\newcommand{\beq}{\begin{equation}}
\newcommand{\eeq}{\end{equation}}
\newcommand{\beqa}{\begin{eqnarray}}
\newcommand{\eeqa}{\end{eqnarray}}
\newcommand{\dv}{\nabla\!\cdot}

\begin{document}
\title{The Stellar-Disk Electric (Short) Circuit: Observational
Predictions for a YSO Jet Flow}

\shorttitle{Stellar Disk Short Circuit}        
\shortauthors{Liffman}

\author{Kurt Liffman \altaffilmark{1}}
\affil{CSIRO/MMT, P.O. Box 56, Highett VIC, Australia 3190, 
 Kurt.Liffman@csiro.au}

\altaffiltext{1}{Department of Mathematical Sciences, 
Monash University, Australia}



\begin{abstract}

We discuss the star-disk electric circuit for a young
stellar object (YSO) and calculate the
expected torques on the star and the disk. We obtain the
same disk magnetic field and star-disk torques as
given by standard magnetohydrodynamic (MHD) analysis.
We show how a short circuit in the star-disk electric
circuit may produce a magnetically-driven
jet flow from the inner edge of a disk surrounding a young star.

An unsteady bipolar jet flow is produced that flows
perpendicular to the disk plane.
Jet speeds of order hundreds of kilometres per second are possible, while
the outflow mass loss rate is proportional to the
mass accretion rate and is
 a function of the disk inner radius relative to the disk co-rotation radius.
\end{abstract}



\section{Star-Disk Magnetic Interaction}
\label{sec:introduction}

\begin{figure}[ht!]
\begin{center}
      \includegraphics[width=0.4\textwidth]{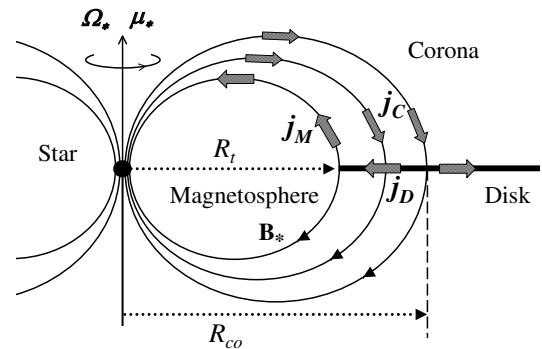} \\
\end{center}
\caption{ Current
flows in the inner section of a star/disk circuit.
$\bbv{j_D}$ -  disk current, $\bbv{j_M}$ - field aligned stellar
magnetosphere current, $\bbv{j_C}$ - field aligned coronal currents
$R_t$ - inner disk truncation radius, $R_{co}$ - co-rotation radius. If the
stellar magnetic field pointed in the opposite direction,
 to that shown in the figure, then the direction
of the current flows would reverse.
}
\label{fig:Star_disk_Circuit_1}
\end{figure}

Most studies of the electro-magnetic interaction between young stars and their
nascent discs undertake their analysis using the
standard magnetohydrodynamic (MHD) approximations (Uzdensky 2004). This approach
has the advantage of describing the Lorentz force in terms
of magnetic fields and allows an analysis that can ignore
electric fields and currents. In this paper, we examine the
star-disk electric circuit to see if we not only obtain the same
answers as standard MHD analysis, but also if we can gain new
insight into how accretion disks and bipolar jets may be related
in young stellar systems.

It is assumed that a star is rotating with an
angular frequency $\bbv{\Omega_*}$, it has a dipole
magnetic field $\bbv{B_*}$, and that the magnetic moment $\bbv{\mu_*}$ is
 aligned with the
rotational axis ($ \hat{\bbv{\mu}}_* =
\hat{\bbv{\Omega}}_*$), which is
perpendicular to the plane of the disk (Fig.~\ref{fig:Star_disk_Circuit_1}).
The direction of $\bbv{B_*}$ is such that the $z$ component
 is negative as it passes through the
accretion disk.

The  stellar magnetic field truncates the disk at a radial distance $R_t$ from
the centre of the star, where
\beqa
\label{eq: R_t}
&&R_t \approx \left(\frac{4\pi}{\mu_0}\frac{B_*^2R_*^6}{\dot{M}_a\sqrt{GM_*}} \right)^{2/7} \
= 0.067 \times \\ 
&& \left( \frac{\left(B_*(R_*)/0.1\ {\rm T}\right)^2\left(R_*/2{\rm R}_\odot\right)^6}
{\left(\dot{M}_a/10^{-8}\ {\rm M}_\odot\ {\rm year}^{-1}\right)
\left( M_*/{\rm M}_\odot\right)^{1/2}} \right)^{2/7} \ {\rm AU} , \nonumber
\eeqa
with $\mu_0$ the permeability of free space,
$\dot{M}_a$ the disk mass accretion rate, $B_*(R_*)$ is the magnetic
field strength at the surface of the star, $R_*$ is the radius of the
star, $M_*$ the stellar mass and $G$
the universal gravitational constant.

 The co-rotation distance, $R_{co}$, is the radial
distance from the star where the
angular frequency of the stellar magnetic field ($\approx \Omega_*$) equals the
Keplerian angular frequency of the disk $\Omega_K(r)$:
\beqa
\label{eq: r_co rotation}
&&R_{co} = \left(\frac{GM_*}{\Omega_*^2}\right)^{1/3} \\
&&= 0.078 \ \left(\left(\frac{M_*}{{\rm M}_\odot}\right)
\left(\frac{P_*}{8 \ {\rm days}}\right)^2\right)^{1/3} \ {\rm AU} \ , \nonumber
\eeqa
with $P_*$ the rotational period of the star. $\Omega_K(r)$
is given by the equation
\begin{equation}
\Omega_K(r)  = \sqrt{\frac{GM_*}{r^3}} \ .
\end{equation}

The relative difference in angular velocity between the
disk and the co-rotating stellar magnetic field generates a current
within the disk. In Fig.~\ref{fig:Star_disk_Circuit_1}, we show
a section of the stellar/disk circuit. Here the current density
generated within
the disk, $\bbv{j_D}$, travels along the inner stellar
magnetic field lines, $\bbv{j_M}$,
and then returns to the disk via the outer stellar magnetic field lines
in the corona above the disk,
$\bbv{j_C}$.
Only the current flow between $R_t$ and $R_{co}$ is shown.
The full star-disk circuit is shown in Bardou \& Heyvaerts (1996).

\section{The Star-Disk Electric Circuit}
\label{sec:star_disk_circuit}

\begin{figure}[ht!]
\begin{center}
      \includegraphics[width=0.4\textwidth]{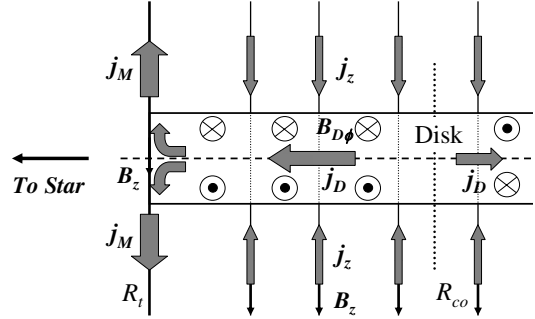}
\end{center}
\caption{ Current flows and magnetic fields near or in the disk. The
poloidal field, $B_z$, is the section of the stellar magnetosphere that
interacts with the disk. The toroidal disk field, $B_{D \phi}$,
and the current flows are generated by the relative motion
between the disk and $B_z$. In this case, $j_D$ is the disk current
density, $j_M$ is the magnetospheric current density that travels
between the inner edge of the disk and the star, while $j_z$ is the
$z$ component of $\bbv{j_C}$: the current between the star and the disk.}
\label{fig:disk_current_and_B_field}
\end{figure}

In Fig.~\ref{fig:disk_current_and_B_field}, we schematically depict
the current flows in and around the inner region of the disk.
The electric field in the disk, $\bbv{E_D}(r)$, is given by (Liffman and Bardou 1999)
\beq
\bbv{E_D}(r) = r\left(\Omega_K(r) - \Omega_*\right)B_{*z}(r)\hat{\bbv{r}}
 ,
\label{eq:disk_Electric_Field}
\eeq
\beq
{\rm with} \ |B_{*z}|(r) \approx B_*(R_*)\left(\frac{R_*}{r}\right)^3 \ ,
\label{eq:B_*z(r)}
\eeq
where, from Fig.~\ref{fig:Star_disk_Circuit_1},
$B_{*z}(r) = -|B_{*z}(r)|$.

For a disk with finite conductivity, $\sigma_D$, the induced electric field
drives a radial current in the disk
with a current density of the form
\beq
\bbv j_D(r) = -\sigma_D(r) r \Omega_K(r)
\left[ \left(\frac{r}{R_{co}} \right)^{3/2} -1  \right]
B_{*z}(r) \ \hat{\bbv r}.
\label{eq:j_Freeman}
\eeq
This radial disk current generates a
toroidal magnetic field in the disk. The equation for which is
(Campbell 1992):
\beq
B_{\phi} = \mu_0 \sigma r z (\Omega_* - \Omega_K(r))B_z(r) \ .
\label{eq:Bphi}
\eeq
Campbell used standard MHD to derive Eqn~(\ref{eq:Bphi}),
but the same result is also obtained from the current flow
model of Fig.~\ref{fig:disk_current_and_B_field} (Liffman \& Bardou 1999).

To compute the field aligned current, $j_z$ $i.e.$, the
$z$ component of $\bbv{j_C}$ that enters the disk
(Fig.~\ref{fig:disk_current_and_B_field}), we apply the steady state
form of conservation of electric charge ($ \dv\bbv j = 0 $), 
which implies
\beq
j_z = \frac{-1}{\beta \mu_0 r}\left[\left( \frac{r}{R_{co}}\right)^{3/2} -
\frac{5}{2} \right]B_{*z}(r) \ ,
\label{eq: j_z}
\eeq
where $\beta$ is a non-dimensional parameter with the definition
(Matt and Pudritz 2005)
\beqa
\label{eq:beta}
&&\beta^{-1} = \mu_0 \sigma_D r h \Omega_K(r) = 2.8 \times \\
&& \left( \frac{\sigma_D}{10^{-7} \ {\rm S m^{-1}}} \right)
\left( \frac{r}{0.1  \ {\rm AU}} \right)
\left( \frac{h}{10^{-3} {\rm AU}} \right)
\left( \frac{\Omega_K}{10^{-5} \ {\rm s}} \right) \ , \nonumber
\eeqa
with $h$ being the scale height of the disk.

We denote by $I_M$ the total current from the top or bottom half of the inner disk
($i.e.$, the inner edge located at the truncation radius, $R_t$) that
travels along the stellar field lines to the star (the corresponding
current density, $j_M$ is shown in Fig.~\ref{fig:Star_disk_Circuit_1}).
the magnitude of $I_M$ is given by (using Eqn (\ref{eq:j_Freeman}))

\beq
I_M = \frac{2\pi R_t}{\mu_0 \beta}
\left| \left(\frac{R_t}{R_{co}}\right)^{3/2} - 1\right|
|B_{*z}(R_t)| \ .
\label{eq: I_MBeta}
\eeq
A representative value for the magnitude of $I_M$
is given by

\beqa
\label{eq:I_M value}
I_M & = & 8.6\times10^{12}
\left(\frac{0.05 \ {\rm AU}}{R_t}\right)^{2}
\left(\frac{B_*(R_*)}{0.1 \ {\rm T}}\right)\times \nonumber\\
& &
\left( \frac{2.8}{\beta} \right)
\left(\frac{R_*}{2{\rm R}_\odot} \right)^3
\left| \left(\frac{R_t}{R_{co}}\right)^{3/2} - 1\right| \ {\rm A} .
\eeqa

\section{Disk-Star Torque}
\label{sec: toroidal torque}

\begin{figure}[ht!]
\begin{center}
      \includegraphics[width=0.4\textwidth]{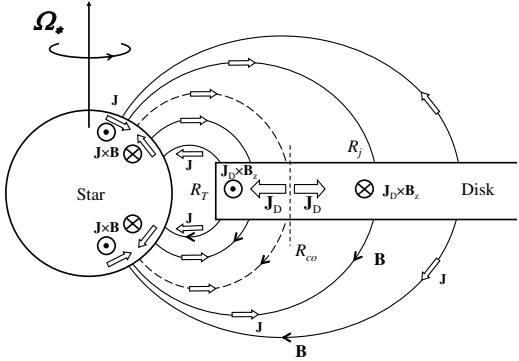} \\
\end{center}
\caption{The interaction of an assumed bipolar
stellar magnetic field, $\bbv{B}$ and the disk produces
current flows, $\bbv{J}$, (denoted by fat arrows) that, in turn,
create $\bbv{J}\times\bbv{B}$ forces which act upon
the star and the disk. For the system shown, the
general rotation is anticlockwise when viewed
from above. The $\bigodot$ and $\bigotimes$ symbols represent the Lorentz force pointing in
the direction towards and away from the observer, respectively.}
\label{fig: disk_star_torque}
\end{figure}

To determine the torque(s) on the disk,
 we consider an annulus of the disk, which has a radius of $r$, thickness $\Delta r$ and
height $2h$. The volume, $\Delta V$, of the annulus is $4 \pi r h \Delta r$ and it feels a torque
\beq
\Delta \tau = \left|\bbv{r}\times\left(\bbv{j_D}\times\bbv{B}_{*z}\right) \right| \Delta V
= 4 \pi r^2 h j_D B_{*z} \Delta r
\label{eq:delta_tau}
\eeq
Substituting Eqns~(\ref{eq:j_Freeman}) and (\ref{eq:beta})
into Eqn~(\ref{eq:delta_tau}) gives the
gradient of the torque exerted by the stellar magnetic field onto the disk:
\beq
\frac{\Delta \bbv{\tau}}{\Delta r} = \frac{4\pi}{\mu_0} r^2 \beta^{-1}
\left[ \left(\frac{r}{R_{co}} \right)^{3/2}- 1\right] B_z(r)^2  \ \hat{\bbv{z}} \ .
\label{eq:D_tau_D_r}
\eeq
Substituting Eqns~(\ref{eq:j_Freeman}) and (\ref{eq:Bphi}) into Eqn~(\ref{eq:delta_tau}) gives
\beq
\left| \frac{\Delta \tau}{\Delta r} \right| = \frac{4 \pi r^2 B_{\phi} B_{*z}}{\mu_0}
\label{eq:Delta_tau_Delta_r}
\eeq
The same equation has been derived via standard MHD analysis (Clarke $et \ al.$ 1995).
This suggests that the equations for the
currents and magnetic fields, as given here, have the correct form. 

\section{The Short Circuit Model}
\label{sec: MPD model}

We now assume  that a portion of the inner field-aligned current, $\bbv{j_M}$, 
(Fig.~\ref{fig:disk_current_and_B_field})
 short circuits  and
produces a radial current.
 A quantitative discussion of such transfield current flows
 is given in Chapters 4 and 7 of Brekke (1997),
where it is shown that transfield currents, such as gravitational drift currents,
regularly occur in the Earth's ionosphere
and magnetosphere.

\begin{figure}[ht!]
\begin{center}
      \includegraphics[width=0.4\textwidth]{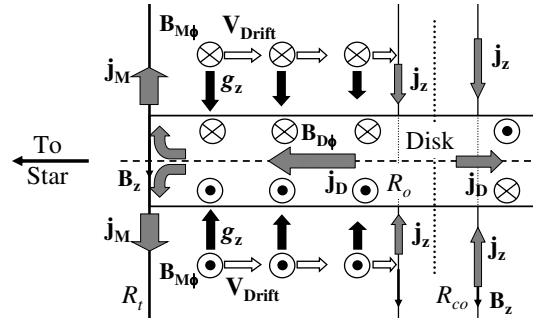}
\end{center}
\caption{ The $z$ component of the stellar gravitational field, $g_z$,
interacts with the toroidal magnetic field above the disk to produce
a radial drift current with a velocity $\bbv{V}_{\rm Drift}$.}
\label{fig:Gravitational_Drift_Current}
\end{figure}

To illustrate how, for example, a radial, gravitational drift current could arise
from the field configuration shown in Fig.~\ref{fig:disk_current_and_B_field},
we note, from Brekke (1997), that the drift velocity, $\bbv{V_D}$,  of a charge, $q$,
subject to a force, $\bbv{F_\perp}$, perpendicular to a magnetic field, $\bbv{B}$, is
given by
\beq
\bbv{V_D} = \frac{\bbv{F_\perp}\times\bbv{B}}{qB^2} \ .
\label{eq:drift}
\eeq 

Using Eqn~(\ref{eq:drift}) we schematically
show, in Fig.~\ref{fig:Gravitational_Drift_Current}, 
how the cross product of the "wound up" toroidal field
($i.e.$, $\bbv{B_{M \phi}}$, produced 
by the disk/star current flow, $I_M$)
and the $z$ component of the stellar gravitational force
can produce a radial, gravitational drift of positively
charged particles (and hence a current) above and below the disk. 
Other particle drifts are also possible, the 
gravitational drift case is shown for purposes of illustration.

It is presumed that this hypothetical `short-circuit' 
region has an inner radius of $ r = R_t$ and an
 outer radius, $R_o$, where $R_o < R_{co}$.
 The radial transfield current, ($\bbv{j_r}$), can interact
with the toroidal field, $\bbv{B_{M \phi}}$.
The subsequent $\bbv{j_r}\times\bbv{B_{M \phi}}$ Lorentz force is in the
correct direction to power a flow away from the disk.

\begin{figure}[ht!]
\begin{center}
      \includegraphics[width=0.4\textwidth]{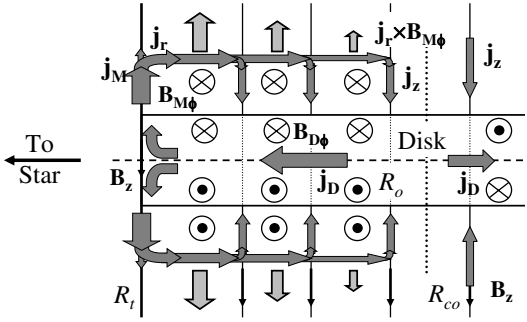}
\end{center}
\caption{ Even in a highly-conductive plasma,
transfield currents can flow between the stellar magnetic field
lines above and below the accretion disk between $R_t$ and $R_o$. The interaction
between the transfield currents and the toroidal fields give rise to
the  $\bbv{j}\times\bbv{B}$ Lorentz forces, which drive the outflow}
\label{fig:Pedersen_and_disk_Currents}
\end{figure}

For the jet flow shown
in Fig.~\ref{fig:Pedersen_and_disk_Currents}, the
flow can only escape the stellar magnetosphere when:
\beq
\frac{1}{2}\rho v^2 \geq \frac{B(R_t)^2}{2\mu_0}
\sim \frac{B_*(R_*)^2}{2\mu_0}\left(\frac{R_*}{R_t}\right)^6 \ ,
\label{eq:energy_break_out}
\eeq
where, $\rho$ is the gas mass density, $v$ the wind speed and it is assumed
that the main part of the flow occurs at the inner edge of the
disk ($r = R_t$).
To find the required values of $\rho$ and $v$, we note
that the velocity of the flow has to be of order the escape speed:
\beq
v \sim \sqrt{\frac{G M_*}{r}} = 133 \
\sqrt{\frac{\left(M_*/{\rm M}_\odot \right)}
{\left( r/0.05 {\rm AU}\right)}} \ {\rm km \ s}^{-1}\ .
\label{eq:v_escape}
\eeq
Combining Eqns~(\ref{eq:energy_break_out}) and (\ref{eq:v_escape})
gives
\beqa
\label{eq:rho_break}
&& \rho \sim \frac{B_*(r)^2 r}{\mu_0 G M_*} 
 = 1.86 \times 10^{-13}\times \\
&&  \frac{\left(B_*(R_*)/0.01 {\rm T}\right)^2
\left(R_*/2{\rm R}_\odot \right)^6}
{\left(M_*/{\rm M}_\odot \right)\left(r/0.05 {\rm AU} \right)^5} \
{\rm kg} \ {\rm m}^{-3}. \nonumber
\eeqa
The existence of such a `break-out'
energy suggests the possibility of a pulsatile jet flow.

\subsection{Jet Exhaust Speed}
\label{sec:exhaust_speed}

 Applying Amperes Law to a thin slice (thickness $dz$)
of a region above the disk
gives
\beq
\frac{\partial B_\phi}{\partial z}(r,z)=-\mu_0 j_r(r,z) \ .
\label{eq:diff_B_phi}
\eeq
As an illustrative example, we will assume a
constant radial current in $z$ for $z \in [z_0,z_T]$,
where $z_0$ and $z_T$ are the bottom and top, respectively,
 of the outflow acceleration region. 
We denote this constant, $z$ independent, radial
current density by $J_{r}$. By also assuming that
$B_\phi(r,z_T) = 0$, we can solve Eqn~(\ref{eq:diff_B_phi})
to obtain
\beq
B_\phi(r,z)=\mu_0 (z_T-z_0)J_r(r)
\left(1 - \frac{z-z_0}{z_T - z_0}\right) \ .
\label{eq:B_phi}
\eeq

The total radial current, $I_r(r)$, is given by
\beq
I_r(r) = 2\pi r(z_T-z_0)J_r(r) \ .
\label{eq:I_r}
\eeq
Combining Eqns~(\ref{eq:B_phi}) and (\ref{eq:I_r}) gives
\beq
B_\phi(r,z)=\frac{\mu_0 I_r(r)}{2\pi r}
\left(1 - \frac{z-z_0}{z_T-z_0}\right)  , \ z \in [z_0,z_T] \ .
\label{eq:B_phi_I_r}
\eeq

Liffman \& Siora (1997) obtained a
Bernoulli equation for the flow:
\beq
 {v^2 \over 2}
 + \left(\frac{\gamma}{\gamma - 1} \right)\frac{p}{\rho}
 +  \frac{B^2}{\mu_0 \rho}
 - \frac{GM_*}{\sqrt{r^2 + z^2}} + \frac{GM_*r_0}{2r^2}
  = \cal{E} ,
\label{eq:Bernoulli3}
\eeq
with $\cal{E}$ the constant specific energy of the streamline flow,
 $\gamma$ - the ratio of specific heats, $B$ is the toroidal magnetic field,
$p$ - pressure,
$r_0$ - the initial value of $r$, and $v = \sqrt{v_r^2 + v_z^2}$.

Using Eqns~(\ref{eq:B_phi_I_r}) and (\ref{eq:Bernoulli3}) one
can obtain an expression for the exhaust speed of the jet flow, $v_e$:

\beqa
\label{eq:v_e}
&&v_e \approx \sqrt{\frac{\mu_0}{2\rho_0}} \
\frac{I_r(r_0)}{\pi r_0}
= 337.3 \times  \\ 
&&\sqrt{\frac{10^{-12} \ {\rm kg} \ {\rm m}^{-3}}{\rho_0}}
\left( \frac{I_r}{10^{13} \ {\rm A}} \right)
\left( \frac{0.05 \ {\rm AU}}{r_0} \right) \ {\rm km} \ {\rm s}^{-1} \ . \nonumber
\eeqa

where $\rho_0$ is the gas density at the base of the flow.
The representative values for $\rho_0$, $I_r$ and $r_0$, as given in
Eqn~(\ref{eq:v_e}), are obtained from Eqns~(\ref{eq:rho_break}), (\ref{eq:I_M value}), 
and (\ref{eq: R_t}), respectively.

\subsection{Radial Size of the Outflow Region}

\label{sec:I_M = integrated j_z}

We denote by $R_o$ the distance from the star, where
 the integrated current density entering
the top half of the disk ($j_z$ - as depicted in Fig.~\ref{fig:Pedersen_and_disk_Currents})
 is equal to the integrated current density returning to the disk
via the stellar magnetosphere
($j_M$ and $j_r$ as in Fig.~\ref{fig:Pedersen_and_disk_Currents}).

Let $I_z(r)$ denote the total vertical current entering the top
half of the disk from the inner edge of the accretion disk, $R_t$, to
a distance $r$ from the star:
\beq
 I_z(r) = \int^r_{R_t}j_z(r)2\pi r dr \ .
 \label{eq:I_z(r)}
 \eeq
 Substituting Eqns~(\ref{eq: j_z}) and (\ref{eq:B_*z(r)}) into Eqn~(\ref{eq:I_z(r)})
 implies
 \beqa
\label{eq:Iz(r)2}
 &&I_z(r) = - \frac{2 \pi R_t B_{*}(R_t)}{\mu_0 \beta} \times  \\
 &&
  \left( 2\left( \frac{R_t}{R_{co}}\right)^{3/2}
 \left[ 1 - \left( \frac{R_t}{r}\right)^{1/2} \right]
 - \frac{5}{4} \left[ 1 - \left( \frac{R_t}{r}\right)^{2} \right]
 \right). \nonumber
 \eeqa

If we consider the case where
 all of the $I_M$ (Eqn~(\ref{eq: I_MBeta})) current short circuits via
the radial transfield current $j_r$ and then back to the disk
via the field-aligned current $j_z$ then, for this case, $R_o$ is determined by
 equating $I_M$ and $I_z(r)$, and specifying that
 $r = R_o$. This condition implies that
 \beqa
\label{eq:Ro_condition}
 \left( \frac{R_t}{R_{co}} \right)^{3/2} &-& 1 =
  2\left( \frac{R_t}{R_{co}}\right)^{3/2}\times \\
 &&\left[ 1 - \left( \frac{R_t}{R_o}\right)^{1/2} \right]
 - \frac{5}{4} \left[ 1 - \left( \frac{R_t}{R_o}\right)^{2} \right] \ .  \nonumber
 \eeqa

\begin{figure}[ht!]
\begin{center}
      \includegraphics[width=0.45\textwidth]{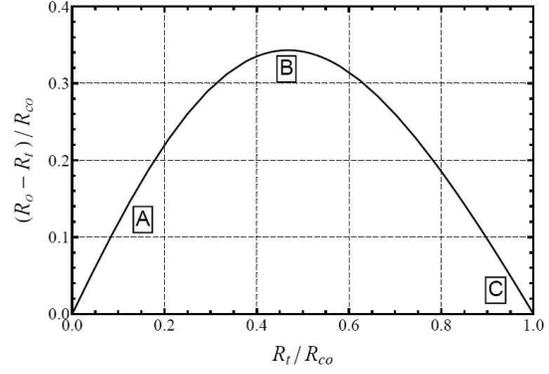}
\end{center}
\caption{The length of the outflow active region of the inner disk ($R_o - R_t$)
as a function of the inner truncation radius, $R_t$, of the disk, where both
quantities are normalized to the co-rotation radius, $R_{co}$. The widths of
the outflow at points
{\bf A}, {\bf B} and {\bf C}
are depicted schematically in Fig.~\ref{fig:mass flow and size}}
\label{fig:RomsRtdRco}
\end{figure}

 We can numerically solve for $R_o$ in the above
 equation and obtain values
 for $(R_o - R_t)/R_{co}$, which is the normalized length of the outflow
 region in the inner accretion disk. These results are shown in 
Fig.~\ref{fig:RomsRtdRco}, where it can be seen that
 as the inner disk approaches the star ($R_t \rightarrow R_*$)
 the width of the outflow acceleration region decreases ($R_o \rightarrow R_t$).
 Similarly, as the inner disk approaches the co-rotation radius the width
 of the outflow acceleration region decreases to zero. The maximum width of
 the acceleration region occurs when the inner disk radius is approximately
 half that of
 the co-rotation radius. This behaviour is represented schematically in
  Fig.~\ref{fig:mass flow and size}, , where we note that the outer radius
  of the outflow acceleration region is always less than the co-rotation
  radius ($R_o < R_{co}$).

\subsection{Mass Ejection Rate}

\begin{figure}[ht!]
\begin{center}
      \includegraphics[width=0.45\textwidth]{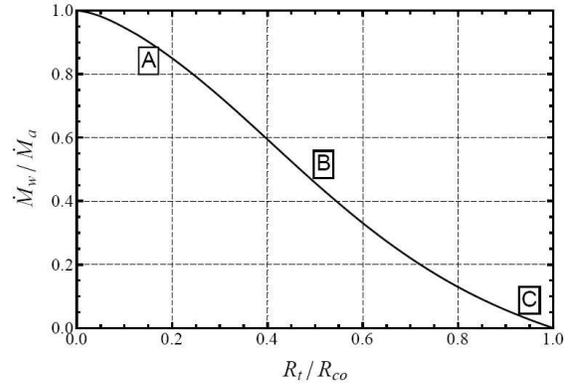}
\end{center}
\caption{The ratio of outflow mass rate, $\dot{M}_w$, to the
mass accretion rate onto a star, $\dot{M}_a$,
versus the ratio of the inner disk truncation radius to the
co-rotation radius ($R_t/R_{co}$). The mass flow rates of
the outflow at points
{\bf A}, {\bf B} and {\bf C}
are depicted schematically by the length of the arrows
in Fig.~\ref{fig:mass flow and size}}
\label{fig:ModM}
\end{figure}

\begin{figure}[ht!]
\begin{center}
      \includegraphics[width=0.45\textwidth]{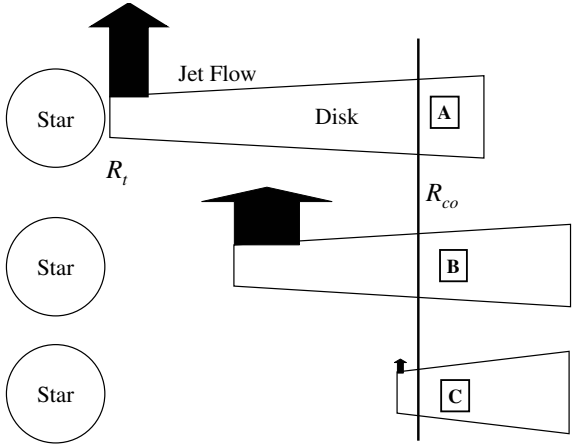}
\end{center}
\caption{A schematic depiction of the mass outflow rate and
the radial size of the outflow acceleration region as a function
of the inner disk truncation radius, $R_t$, and the co-rotation radius,
$R_{co}$ - indicated by the line.
The length of the arrow represents the outflow mass rate, $\dot{M}_w$,
while the width of the arrow indicates the actual, relative size of
the outflow acceleration region. Case {\bf A}: $R_t$ is small
relative to $R_{co}$ and $ \dot{M}_w \sim \dot{M}_a$. {\bf B}:
 $R_t \sim R_{co}/2$, the outflow acceleration region is at its
 broadest and $ \dot{M}_w \sim \dot{M}_a/2$. {\bf C}: $R_t \sim R_{co}$
 and $ \dot{M}_w \rightarrow 0$.}
\label{fig:mass flow and size}
\end{figure}

From the conservation of mass, the mass ejection rate of an outflow,
$ \dot{M}_w$, is
\beq
\dot{M}_w = \rho v A \ ,
\label{eq:Mdot}
\eeq
where $\rho$, $v$ and $A$ are, respectively,
the density, speed and cross-sectional
 area of the outflow. Noting that
 \beq
 A = \pi(R_o^2 - R_t^2)
 \eeq
 and using Eqns~(\ref{eq:v_escape}), (\ref{eq:rho_break})
 and (\ref{eq: R_t}), Eqn~(\ref{eq:Mdot})
 has the form
 \beq
 \dot{M}_w
 = \frac{\dot{M}_a}{4}\left[ \left( \frac{R_o}{R_t}\right)^2 - 1 \right]
 \label{eq:dotM_w}
 \eeq

Using Eqns~(\ref{eq:dotM_w}) and (\ref{eq:Ro_condition}), we can compute the ratio $\dot{M}_w/\dot{M}_a$
as a function of $R_t/R_{co}$. These results are shown in Fig.~\ref{fig:ModM}, where
it can be seen that when $R_t = R_{co}$ the high-speed outflow shuts down. On
the other hand, for a fixed co-rotation radius, the mass outflow rate increases and approaches the mass accretion rate as the inner disk radius approaches the
surface of the star. This behaviour is represented schematically in
  Fig.~\ref{fig:mass flow and size}.

The observed values for the mass outflow and accretion rates are only
known to order-of-magnitude values: $\dot{M}_w/\dot{M}_a \sim 0.1$ (Calvet 1997),
while the modeling of T Tauri observational data gives:
$R_t/R_{co} \approx 0.6 \ {\rm to} \ 0.8 \ $ (Kenyon $et \ al.$ 1996).
From Fig.~\ref{fig:ModM}, the corresponding range of
values for $\dot{M}_w/\dot{M}_a$ is
$\dot{M}_w/\dot{M}_a \approx 0.33 \ {\rm to} \ 0.13 \ $.
These values are consistent with the observational values.

\section{Conclusions}

The star-disk electric circuit arises due to the
interaction of the stellar magnetosphere and the accretion disk.
 We have shown that the electric circuit model and standard MHD analysis give the
 same expressions for the disk magnetic field and
the torque between the disk and the star.

We examined the hypothetical case where there is a short circuit in
the star-disk circuit. This radial short circuit
may be due to a gravitational drift current, which might be of sufficient magnitude
to generate a non-constant
bipolar outflow at the inner edge of the disk that flows in a
direction roughly perpendicular to the disk.

In this model, the irregularity in the outflow arises
because the outflow has to disrupt
the stellar magnetic field for it to escape the stellar-disk system.
The model predicts that the mass outflow rate is proportional to the
total mass accretion rate in the disk.
The mass outflow rate is also dependent on the position of the
inner edge of the disk relative to the disk co-rotation radius. If the
position of the inner edge of the disk is equal to the
disk co-rotation radius, then there
is little or no outflow. As the inner edge of the disk approaches the
star (assuming a constant co-rotation radius)
then the proportion of material going into the outflow
increases, while the proportion of material accreting onto the star
decreases.

\end{document}